\documentclass[pra,prl,twocolumn,showpacs,floatfix,10pt]{revtex4-1}
\usepackage{graphicx}% Include figure files
\usepackage{dcolumn}% Align table columns on decimal point
\usepackage{bm}% bold math
\usepackage{subfigure}
\usepackage{float}
\usepackage{multirow}
\usepackage{booktabs}
\usepackage{amsmath}
\usepackage{latexsym,amssymb}
\usepackage[mathscr]{eucal}
\usepackage[bookmarks=true,
   colorlinks=true,
   linkcolor=blue,
   urlcolor=blue,
   citecolor=blue,
   bookmarks=true,
   hyperindex=true
]{hyperref}

\begin{document}

%\preprint{APS/123-QED}

\title{Thermodynamic phase transition of a black hole in rainbow gravity\\}

\author{Zhong-Wen Feng\textsuperscript{1,2,}}
\altaffiliation{Email: zwfengphy@163.com}
\author{Shu-Zheng Yang\textsuperscript{1,2}}
\altaffiliation{Email: szyangcwnu@126.com}
\vskip 0.5cm
\affiliation{1 Physics and Space Science College, China West Normal University, Nanchong, 637009, China\\
2 Department of Astronomy, China West Normal University, Nanchong 637009, China}

\date{\today}% It is always \today, today,

\begin{abstract}
In this letter, using the rainbow functions that were proposed by Magueijo and Smolin, we investigate the thermodynamics and the phase transition of rainbow Schwarzschild black hole. First, we calculate the rainbow gravity corrected Hawking temperature. From this modification, we then derive the local temperature, free energy, and other thermodynamic quantities in an isothermal cavity. Finally, we analyze the critical behavior, thermodynamic stability, and phase transition of the rainbow Schwarzschild black hole. The results show that the rainbow gravity can stop the Hawking radiation in the final stages of black holes' evolution and lead to the remnants of black holes. Furthermore, one can observe that the rainbow Schwarzschild  black hole has one first-order phase transition, two second-order phase transitions, and three Hawking-Page-type phase transitions in the framework of rainbow gravity theory.
\end{abstract}

%\pacs{04.60.Bc, 04.80.Cc, 03.75.Dg}% PACS, the Physics and Astronomy
\maketitle

\section{Introduction}
\label{Int}
The Lorentz symmetry is known as one of the fundamental symmetries in nature, which produces the standard energy-momentum dispersion relation, i.e.,  $E^2  - p^2  = m^2$\cite{ch1,ch2}. However, many works claimed that the standard energy-momentum dispersion relation would not be held when it nears the Planck length  (or the Planck energy) \cite{ch3,ch4,ch5,ch6,ch7}. In fact, one of the intriguing predictions among various quantum gravity theories, such as loop quantum gravity, loop quantum gravity, and non-commutative geometry, is the existence of a minimum measurable length that can be identified with the Planck length \cite{ch8,ch9,ch10}. This idea is supported by many Gedanken experiments \cite{ch11}. Furthermore, the Planck length can be considered as the dividing line  between the quantum and the classical description of spacetime. Therefore, the Planck length should be taken as an invariant scale. However, in the special relativity, these features conflict with the Lorentz symmetry because the Planck scale is not invariant under the linear Lorentz transformations. To solve this paradoxical situation, the standard energy-momentum dispersion relation must be changed to the so-called modified dispersion relation (MDR).

Combining the MDR with the special relativity, Amelino-Camelia proposed the double special relativity (DSR) \cite{ch21}. The DSR is an extension of special relativity, which contains two fundamental constants: the velocity of light  $c$ and the Planck energy $E_p$. Meanwhile, the DSR is also a framework for encoding properties of flat quantum spacetime \cite{ch22}. However, it is observed that the DSR is typically formulated in momentum space. Therefore, a nonlinearity of the Lorentz transformation has a very significant effect on the definition of a dual space in the framework of DSR, i.e., the dual space is non-trivial \cite{ch31}. To overcome this problem, the DSR was generalized to the curved spacetimes by Magueijo and Smolin. This doubly general theory of relativity is called as rainbow gravity (or gravity's rainbow) \cite{ch23}. The name rainbow gravity (RG) comes from the fact that this theory assumes the spacetime background depending on the energy of a test particle. Therefore, one should use a family of metrics (rainbow metrics) parameterized by the ratio ${E \mathord{\left/{\vphantom {E {E_p }}} \right.\kern-\nulldelimiterspace} {E_p }}$  to describe the background of this spacetime instead of using a single metric.

The rainbow gravity is very important because it can be applied to different physical systems and modifies many classical theories. For example, by incorporating the RG with the Friedmann-Robertson-Walker cosmologies, the possibility of resolving big bang singularity was investigated in~\cite{ch24}. In~\cite{ch26}, with the help of the RG, the authors investigated the modified Starobinsky model and the inflationary solution to the motion equations, and the spectral index of curvature perturbation and the tensor-to-scalar ratio were also calculated. In addition, the deflection of light, photon time delay, gravitational red-shift, and the weak equivalence principle were also studied in the framework of the RG \cite{ch33}. It is worth mentioning that the RG has a great effect on the thermodynamics of black holes \cite{ch31,ch32,ch34,ch35,ch36,ch38,ch39,ch40,ch41}. In 2004, Ali \emph{et al.} showed that the RG can prevent the black holes from total evaporation, which leads to remnants of black holes in exactly the same way as done by the generalized uncertainty principle \cite{ch31,ch41+,ch43}. Hence, the RG may solve the information loss and naked singularity problems of black holes.

On the other hand, it is well known that black holes have not only thermodynamic properties, but also rich phase structures and critical phenomena. Over the past decades, the phase transitions and critical phenomena of black holes have been widely explored \cite{ch43+}. In 1977, Davies found that the Kerr-Newman black hole exhibits the phenomena of phase transition \cite{ch44}. Later, Pav\'{o}n discovered a non-equilibrium second order phase transition in the charged Reissner-Nordstr\"{o}m (RN) black hole \cite{ch45}. The existence of a certain phase transition in the asymptotically anti de Sitter (AdS) spacetimes was proved by Hawking and Page. They demonstrated that the AdS Schwarzschild (SC) black hole undergoes a phase transition (i.e., Hawking-Page phase transition) to a thermal AdS space if the temperature reaches a certain value \cite{ch46}. This seminal work has attracted wide attention because it can explained the confinement/deconfinement phase transition of gauge field in the AdS/CFT correspondence \cite{ch47}. Since then, several works were devoted to investigate the Hawking-Page phase transition and the critical phenomena for other more complicated AdS spacetimes \cite{ch50,ch51,ch53,ch54,ch55,ch56,ch57}. From these studies, people found that the phase transitions behavior of charged black holes is similar to that of the Van der Waals liquid-gas system \cite{ch50,ch51,ch53,ch54,ch55,ch56,ch57,ch56+}. In 2012, Kubiz\v{n}\'{a}k and Mann proposed a remarkable new perspective on the relationship between the phase transition of charged AdS black hole and the Van der Waals liquid-gas phase transition \cite{ch50}. By treating the cosmological constant $\lambda$ as a  thermodynamic pressure $P$ and its conjugate variable as specific volume $V$, they analyzed the thermodynamic behavior of RN-AdS black hole in the extended phase space. The results showed that the RN-AdS black hole system has a first-order small-large black hole phase transition since the free energy $G$ demonstrates a ``swallow tail''-type behavior. In addition, Kubiz\v{n}\'{a}k and Mann also studied the ``$P-V$''  criticality and the critical exponents and proved that they coincide with those of the Van der Waals system.

From the above discussion, it is interesting to investigate the thermodynamic criticality and the phase transition of the rainbow black holes. Therefore, in this letter, we study the thermodynamic criticality and phase transition of rainbow SC black hole. By utilizing the relation $E \ge {1 \mathord{\left/ {\vphantom {1 {r_H }}} \right. \kern-\nulldelimiterspace} {r_H }} = {1 \mathord{\left/ {\vphantom {1 {2GM}}} \right. \kern-\nulldelimiterspace} {2GM}}$ and a new kind of rainbow functions, we derive the modified thermodynamic quantities and free energy of the rainbow SC black hole. According to these modifications, the thermodynamic stability, thermodynamic criticality and phase transition of the rainbow SC black hole are analyzed. It turns out that our results are different from those of the standard Hawking-Page phase transition. However, when the ratio ${E \mathord{\left/{\vphantom {E {E_p }}} \right.\kern-\nulldelimiterspace} {E_p }}$ approaches zero, our results reduce to the standard forms.

The organization of this paper is as follows. The next section is devoted to introducing a new kind of rainbow functions, which were proposed by Magueijo and Smolin. In Section~\ref{TP I}, using the rainbow functions, we study the thermodynamics of SC black hole in the context of rainbow gravity, and then its thermodynamic stability, thermodynamic criticality and phase transition are discussed. The paper ends with conclusions in Section~\ref{Dis}.

\section{A brief introduction on the rainbow functions}
\label{RF}
To begin with, we briefly review the MDR and the rainbow functions. The general form of MDR can be expressed as follows:
\begin{align}
\label{eq1}
E^2 f^2 \left( {{E \mathord{\left/ {\vphantom {E {E_p }}} \right. \kern-\nulldelimiterspace} {E_p }}} \right) - p^2 g^2 \left( {{E \mathord{\left/
 {\vphantom {E {E_p }}} \right. \kern-\nulldelimiterspace} {E_p }}} \right) = m^2 ,
\end{align}
where  $E_p$ is the Planck energy, and the correction terms  $f\left( {{E \mathord{\left/{\vphantom {E {E_p }}} \right. \kern-\nulldelimiterspace} {E_p }}} \right) $ and $g\left( {{E \mathord{\left/ {\vphantom {E {E_p }}} \right. \kern-\nulldelimiterspace} {E_p }}} \right)$  are known as rainbow functions, which are responsible for the modification of the energy-momentum relation in the ultraviolet regime. However, in the limit  ${E \mathord{\left/ {\vphantom {E {E_p }}} \right. \kern-\nulldelimiterspace} {E_p }} \to 0$, the rainbow functions satisfy the  following relations
\begin{align}
\label{eq2}
\mathop {\lim }\limits_{{E \mathord{\left/ {\vphantom {E {E_p }}} \right. \kern-\nulldelimiterspace} {E_p }} \to 0} f\left( {{E \mathord{\left/ {\vphantom {E {E_p }}} \right. \kern-\nulldelimiterspace} {E_p }}} \right) = 1,
\mathop {\lim }\limits_{{E \mathord{\left/ {\vphantom {E {E_p }}} \right. \kern-\nulldelimiterspace} {E_p }} \to 0} g\left( {{E \mathord{\left/
 {\vphantom {E {E_p }}} \right. \kern-\nulldelimiterspace} {E_p }}} \right) = 1.
\end{align}
Eq.~(\ref{eq2}) indicates that the MDR reduces to the standard  energy-momentum dispersion relation at low energy scale.
It should be mentioned that the expression of rainbow function is not unique, and people can find a series expressions for rainbow functions depending on different phenomenological motivations. According to the varying speed of light theory, Magueijo and Smolin developed a MDR,
 which takes the form of ${{E^2 } \mathord{\left/  {\vphantom {{E^2 } {\left( {1 - {{\gamma E} \mathord{\left/ {\vphantom {{\gamma E} {E_p }}} \right. \kern-\nulldelimiterspace} {E_p }}} \right)^2 }}} \right.  \kern-\nulldelimiterspace} {\left( {1 - {{\gamma E} \mathord{\left/ {\vphantom {{\gamma E} {E_p }}} \right. \kern-\nulldelimiterspace} {E_p }}} \right)^2 }} - p^2  = m^2$ \cite{ch13}. This MDR implies that a spacetime has an energy-dependent velocity $c = \left( {1 - {{\gamma E} \mathord{\left/ {\vphantom {{\gamma E} {E_p }}} \right. \kern-\nulldelimiterspace} {E_p }}} \right){}$. Comparing the MDR with Eq.~(\ref{eq1}), the rainbow functions can be fixed as follows:
\begin{equation}
\label{eq3}
f\left( {{E \mathord{\left/
 {\vphantom {E {E_p }}} \right.
 \kern-\nulldelimiterspace} {E_p }}} \right) = \frac{1}{{1 - {{\gamma E} \mathord{\left/
 {\vphantom {{\gamma E} {E_p }}} \right.
 \kern-\nulldelimiterspace} {E_p }}}}, \quad  g\left( {{E \mathord{\left/
 {\vphantom {E {E_p }}} \right.
 \kern-\nulldelimiterspace} {E_p }}} \right) = 1,
\end{equation}
where $\gamma$ is the rainbow parameter. Eq.~(\ref{eq3}) implies that the varying velocity of light in the RG becomes smaller when the energy of photons increases. Notably, the bounds on the values of  $\gamma$ can be analyzed by using many theoretical and experimental considerations \cite{ch33,ch58+}. The concrete expressions of rainbow functions have a strong effect on the predictions. Hence, in the subsequent discussions, we scrutinize the thermodynamics, phase structures, and critical phenomena of rainbow SC black hole using rainbow functions~(\ref{eq3}).

\section{Thermodynamics and phase transition of rainbow SC black hole}
\label{TP I}
\begin{figure}
\centering
\subfigure[]{
\begin{minipage}[]{0.44\textwidth}
\includegraphics[width=1.1\textwidth]{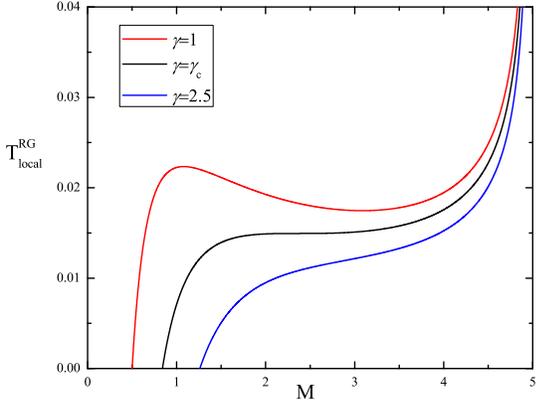}
\label{fig1-a}
\end{minipage}
}
\subfigure[]{
\begin{minipage}[]{0.44\textwidth}
\includegraphics[width=1.1\textwidth]{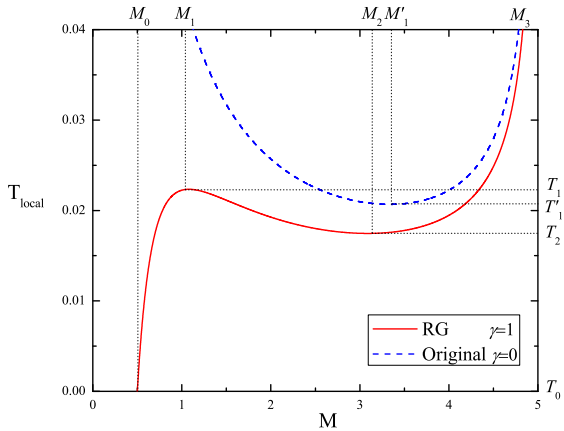}
\label{fig1-b}
\end{minipage}
}
\caption{(a) Relationship between the rainbow local temperature and mass for different $\gamma$.
(b) Original and modified local temperature versus mass. We set $ G = 1$ and  $r = 10$.}
\label{fig1}
\end{figure}

In this section, we investigate the thermodynamics and phase transition of rainbow SC black hole using rainbow functions~(\ref{eq3}). In \cite{ch23,ch43},
 the authors pointed out that one can construct rainbow spacetimes by replacing $dt \to {{dt} \mathord{\left/ {\vphantom {{dt} {f\left( {{E \mathord{\left/ {\vphantom {E {E_p }}} \right. \kern-\nulldelimiterspace} {E_p }}} \right)}}} \right. \kern-\nulldelimiterspace} {f\left( {{E \mathord{\left/ {\vphantom {E {E_p }}} \right. \kern-\nulldelimiterspace} {E_p }}} \right)}}$  for time coordinates and $dx^i  \to {{dx^i } \mathord{\left/ {\vphantom {{dx^i } {g\left( {{E \mathord{\left/ {\vphantom {E {E_p }}} \right. \kern-\nulldelimiterspace} {E_p }}} \right)}}} \right. \kern-\nulldelimiterspace} {g\left( {{E \mathord{\left/ {\vphantom {E {E_p }}} \right. \kern-\nulldelimiterspace} {E_p }}} \right)}}$  for all spatial coordinates. Therefore, the line element of rainbow SC black hole is given as follows:
\begin{align}
\label{eq4}
ds^2 & = - \frac{{1 - \left( {{{2GM} \mathord{\left/  {\vphantom {{2GM} r}} \right.  \kern-\nulldelimiterspace} r}} \right)}}{{f^2 \left( {{E \mathord{\left/
 {\vphantom {E {E_p }}} \right.  \kern-\nulldelimiterspace} {E_p }}} \right)}}dt^2  + \frac{{dr^2 }}{{\left[ {1 - \left( {{{2GM} \mathord{\left/
 {\vphantom {{2GM} r}} \right.  \kern-\nulldelimiterspace} r}} \right)} \right]g^2 \left( {{E \mathord{\left/
 {\vphantom {E {E_p }}} \right.  \kern-\nulldelimiterspace} {E_p }}} \right)}}
\nonumber \\
& +  \frac{{r^2 }}{{g^2 \left( {{E \mathord{\left/
 {\vphantom {E {E_p }}} \right.  \kern-\nulldelimiterspace} {E_p }}} \right)}}d\Omega ^2 ,
\end{align}
where $d\Omega ^2  = d\theta ^2  + \sin ^2 \theta d\phi ^2$  represents the line element of two-dimensional hypersurfaces. Obviously, the event horizon of rainbow SC black hole is located at  $r_H  = 2GM$. When considering Eq.~(\ref{eq2}), the metric of original SC black hole is recovered. According to Eq.~(\ref{eq3}) and Eq.~(\ref{eq4}), the Hawking temperature of rainbow SC black hole can be derived from the definition of surface gravity as follows: \cite{ch62,ch63}
\begin{align}
\label{eq5}
T_H^{{\rm{RG}}}  = \frac{\kappa }{{2\pi }} = \frac{{g\left( {{E \mathord{\left/
 {\vphantom {E {E_p }}} \right.
 \kern-\nulldelimiterspace} {E_p }}} \right)}}{{f\left( {{E \mathord{\left/
 {\vphantom {E {E_p }}} \right.
 \kern-\nulldelimiterspace} {E_p }}} \right)}}T_H  = \frac{1}{{8\pi GM}}\left( {1 - \gamma \frac{E}{{E_p }}} \right),
\end{align}
where $\kappa  =  - \frac{1}{2}\mathop {\lim }\limits_{r \to r_H } \sqrt { - \frac{{g^{11} }}{{g^{00} }}} \frac{{\left( {g^{00} } \right)^\prime  }}{{g^{00} }}  = \frac{{\kappa _0 g\left( {{E \mathord{\left/ {\vphantom {E {E_p }}} \right. \kern-\nulldelimiterspace} {E_p }}} \right)}}{{2\pi f\left( {{E \mathord{\left/
 {\vphantom {E {E_p }}} \right. \kern-\nulldelimiterspace} {E_p }}} \right)}}$ is  the surface gravity. $\kappa _0  = {1 \mathord{\left/ {\vphantom {1 {4GM}}} \right. \kern-\nulldelimiterspace} {4GM}}$ and $T_H  = {1 \mathord{\left/ {\vphantom {1 {8\pi GM}}} \right.\kern-\nulldelimiterspace} {8\pi GM}} $ are the original surface gravity and Hawking temperature of SC black hole, respectively. Obviously, the original surface gravity gets modified in RG. Meanwhile, it is evident that the rainbow Hawking temperature is very sensitive to the concrete expression of rainbow functions. Following the heuristic argument in \cite{ch64,ch65}, the Heisenberg uncertainty principle (HUP)  $\Delta x\Delta p \ge 1$  still holds in RG. Therefore the HUP can be translated into a lower bound on the energy, i.e., $E \ge {1 \mathord{\left/ {\vphantom {1 {\Delta x}}} \right. \kern-\nulldelimiterspace} {\Delta x}}$, where $E$ is the energy of a particle emitted in the Hawking radiation. From the lower bound on the energy and the uncertainty position $\Delta x$, one can obtian the following relation: \cite{ch31}
\begin{equation}
\label{eq7}
E \ge {1 \mathord{\left/
 {\vphantom {1 {\Delta x}}} \right.
 \kern-\nulldelimiterspace} {\Delta x}} \approx {1 \mathord{\left/
 {\vphantom {1 {r_H }}} \right.
 \kern-\nulldelimiterspace} {r_H }} = {1 \mathord{\left/
 {\vphantom {1 {2GM}}} \right.
 \kern-\nulldelimiterspace} {2GM}}.
\end{equation}
It should be noted that this bound on the energy plays a key role in the modification of thermodynamics of rainbow black holes. Substituting Eq.~(\ref{eq7}) into Eq.~(\ref{eq5}), the rainbow Hawking temperature can be rewritten as
\begin{equation}
\label{eq8}
T_H ^{\rm{RG}} = \frac{1}{{8\pi GM}}\left( {1 - \frac{{\gamma  }}{{2 \sqrt G M}}} \right),
\end{equation}
where we use $E_p  = G^{{{ - 1} \mathord{\left/ {\vphantom {{ - 1} 2}} \right. \kern-\nulldelimiterspace} 2}}$ in natural units. When  $\gamma=0$, the rainbow Hawking temperature reaches as that in the original case. Next, we investigate the entropy of rainbow SC black hole. Using the first law of black hole thermodynamics, the entropy of rainbow SC black hole is given by
\begin{align}
\label{eq9}
S& =\int {T_H^{ - 1} dM }
\nonumber \\
&= 4\sqrt G M\pi \left( {\sqrt G M + \gamma } \right) + 2\pi \gamma ^2 {\rm{ln}}\left( {2\sqrt G M - \gamma } \right).
\end{align}
It is clear that the area law of the entropy $S = 4\pi M^2$ is recovered when $\gamma=0$. Since the evolution and the phase transition of black holes can be detected at a finite surface $r$, we need to calculate the local temperature of rainbow SC black hole. On the basis of the finding from~ \cite{ch66,ch59+,ch59}, the local temperature in the RG at a finite distance  $r$ outside the black hole can be expressed as follows:
\begin{equation}
\label{eq9+}
T_{{\rm{\rm{local}}}}^{\rm{RG}}= T_H \left( {1 - \frac{{2GM}}{r}} \right)^{ - \frac{1}{2}} \frac{{g\left( {{E \mathord{\left/
 {\vphantom {E {E_p }}} \right.
 \kern-\nulldelimiterspace} {E_p }}} \right)}}{{f\left( {{E \mathord{\left/
 {\vphantom {E {E_p }}} \right.
 \kern-\nulldelimiterspace} {E_p }}} \right)}}.
 \end{equation}
Considering Eq.~(\ref{eq7}), Eq.~(\ref{eq8}), and the rainbow functions (\ref{eq3}), the corresponding local temperature for the observer on the cavity is given by
\begin{equation}
\label{eq10}
T_{{\rm{\rm{local}}}}^{\rm{RG}} = \frac{1}{{8\pi GM\sqrt {1 - \frac{{2GM}}{r}} }}\left( {1 - \frac{\gamma }{{2\sqrt G M}}} \right).
\end{equation}
Obviously, Eq.~(\ref{eq10}) is implemented by the redshift factor of the metric. If one considers the parameter $r$ as an invariable quantity, the critical value of black hole's mass, the rainbow parameter, and the local temperature can be obtained by the following equations
\begin{equation}
\label{eq10+}
\left( {\frac{{\partial T_{{\rm{\rm{local}}}} }}{{\partial M}}} \right)_r  = 0, \quad   \left( {\frac{{\partial ^2 T_{{\rm{\rm{local}}}} }}{{\partial M^2 }}} \right)_r  = 0.
\end{equation}
Using Eq.~(\ref{eq10+}), and setting $r=10$ and $G=1$, the critical mass, the critical free parameter, and the critical local temperature are
\begin{align}
\label{eq11a+}
M_{c}^{\rm RG}=2.36701,  \gamma_{c}=1.68082,  T_{c}^{\rm RG}=0.0149398.
\end{align}
The local temperature as a function of $\gamma$ is depicted in Fig.~\ref{fig1}.

 From Fig.~\ref{fig1-a}, it can be observed that for $\gamma \neq 0$, there is a phase transition with $0 < \gamma < \gamma_{c}$. Therefore, for the convenience of discussing the critical phenomena and phase transition of rainbow SC black hole, we set $\gamma=1$ in the following discussions.

 By fixing $G = 1$ and $r=10$, we plot the rainbow local temperature $T_{\rm{local}}^{\rm{RG}}$ and  original local temperature $T_{\rm{local}}$ versus the mass of SC black hole in Fig.~\ref{fig1-b}. The blue dashed line illustrates the original local temperature, which is infinite when the mass approaches zero and $M_3$. The minimum value of local temperature $T_1'$  occurs at  $M_1'$. For the rainbow local temperature case (red solid line), one can see that the behavior of $T_{\rm{local}}^{\rm{RG}}$ is similar to that of the original local temperature at  $M_3$, where the horizon of black hole is close to $r = 2GM_3$. This indicates that the black hole is very hot near the event horizon. As the mass of the black hole decreases, the rainbow local temperature reduces to $T_2$ corresponding to the mass $M_2$, and then increases to its maximum value $T_1$ at $M_1$. Finally, the rainbow local temperature becomes zero when the mass of black hole approaches a finite value $M_0$, which leads to a remnant of rainbow SC black hole, i.e., $M_{\rm{res}} = M_0 ={{\gamma \sqrt G } \mathord{\left/ {\vphantom {{\gamma \sqrt G } 2}} \right. \kern-\nulldelimiterspace} 2}$.
The values of $\left( {M_1 ,T_1 } \right)$, $\left( {M_2 ,T_2 } \right)$, $\left( {M_3 ,T_3 } \right)$ and $\left( {M_1' ,T_1' } \right)$ can be numerically obtained if needed.

\begin{figure}
\centering % \begin{center}/\end{center} takes some additional vertical space
\includegraphics[width=.5\textwidth,origin=c,angle=0]{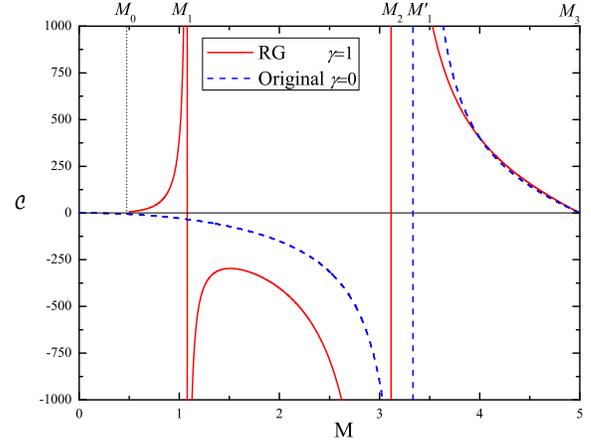}
% "\includegraphics" is very powerful; the graphicx package is already loaded
\caption{\label{fig2} Original and rainbow heat capacity versus mass for $G = 1$ and  $r = 10$.}
\end{figure}

Next, using the first law of thermodynamics, the local energy of rainbow SC black hole within the boundary $r$ is given as follows:
\begin{align}
\label{eq11}
E_{\rm{local}}^{\rm{RG}}& = \int_{M_0 }^M {T_{\rm{local}}^{\rm{RG}}dS^{\rm{RG}}}
\nonumber \\
& = \frac{r}{G}\left( {\sqrt {1 - \frac{{\gamma G^{{3 \mathord{\left/
 {\vphantom {3 2}} \right.
 \kern-\nulldelimiterspace} 2}} }}{r}}  - \sqrt {1 - \frac{{2GM}}{r}} } \right).
\end{align}
When $\gamma \rightarrow 0$, Eq.~(\ref{eq11}) reduces to the original local energy $E_{\rm{local}}  = {{r\left( {1 - \sqrt {1 - {{2GM} \mathord{\left/ {\vphantom {{2GM} r}} \right.  \kern-\nulldelimiterspace} r}} } \right)} \mathord{\left/ {\vphantom {{r\left( {1 - \sqrt {1 - {{2GM} \mathord{\left/ {\vphantom {{2GM} r}} \right.  \kern-\nulldelimiterspace} r}} } \right)} G}} \right. \kern-\nulldelimiterspace} G}$. It may be noted that one can investigate the thermodynamic stability of the black holes from their heat capacity. Hence, using the rainbow local temperature Eq.~(\ref{eq10}) and the rainbow local energy Eq.~(\ref{eq11}), the rainbow heat capacity at fixed $r$ can be expressed as
\begin{align}
\label{eq12}
\mathcal{C}^{\rm{RG}}& = \left( {\frac{{\partial E_{{\rm{local}}}^{{\rm{RG}}} }}{{\partial T_{{\rm{local}}}^{{\rm{RG}}} }}} \right)_r
\nonumber \\
&= \frac{{16M^3 \pi \left( {r - 2GM} \right)}}{{2M\left( {3GM - r} \right) + \gamma \sqrt G \left( {2r - 5GM} \right)}}.
\end{align}
By setting $\mathcal{C}^{\rm{RG}} =0$, the remnant mass is $M_{{\rm{res}}}  = {{\gamma \sqrt G } \mathord{\left/ {\vphantom {{\gamma \sqrt G } 2}} \right.
 \kern-\nulldelimiterspace} 2}$. Obviously, Eq.~(\ref{eq12}) reduces to the original heat capacity $ \mathcal{C}  = 8\pi GM^2 (r - 2GM)/(3GM - r)$ when  $\gamma$ vanishes. The heat capacity associated with the mass of SC black hole is plotted in Fig.~\ref{fig2}.

In Fig.~\ref{fig2}, the red solid line for rainbow heat capacity $\mathcal{C}^{\rm{RG}}$ vanishes at $M_0$, which is different from the blue dashed line for the original heat capacity $\mathcal{C}$ that reaches zero when $M \to 0$. It is well known that the black hole is stable if its heat capacity is positive, while the black hole is unstable if it has negative heat capacity. Therefore, one can find that the rainbow heat capacity has two stable regions $M_0 \leq M \leq M_1$ and  $M_2 \leq M \leq M_3$, and one unstable region $M_1 \leq M \leq M_2$. However, the blue dashed line for the original case has only one stable region $M_1' \leq M \leq M_3$, and one unstable region $ M \leq M_1'$. Interestingly, the rainbow heat capacity diverges at the points where the temperature reaches the maximum value $M_1$ and the minimum value $M_2$, which indicates that two second-order phase transitions  exist in the canonical ensemble.

According to the discussions on the local temperature and the heat capacity, we classify the rainbow SC black hole into three branches depending on its mass scale. The range, state, and stability of the three branches of rainbow SC black hole are presented in Table.~\ref{tab1}.
\begin{table}
\centering
\caption {\label{tab1} Region, state, and stability of the three branches of rainbow SC black hole.}
\begin{tabular}{c c c c c c}
\hline
Branches   &         Region         &  State   & Stability \\
\hline
1         &$ M_0 \leq M \leq M_1$ &   small        &  stable   \\
2         &$ M_1 \leq M \leq M_2$ &  intermediate  &  unstable  \\
3         &$ M_2 \leq M \leq M_3$ &   large        &  stable    \\
\hline
\end{tabular}
\end{table}
From Table.~\ref{tab1}, an additional intermediate black hole (IBH) can be found in the system, which never appears in the original case. Obviously, the stability of the three black holes depend on their heat capacities. Therefore, it is obvious that the small black hole (SBH) and the large black hole (LBH) are stable, showing that they can survive for a long time within the frame of the RG theory. On the contrary, the IBH is unstable because its heat capacity is negative, which implies that it would decay into the SBH or LBH quickly. In addition, it should be noted that the stability of the black holes can be known from their free energy.

\begin{figure}
\centering
\subfigure[]{
\begin{minipage}[b]{0.41\textwidth}
\includegraphics[width=1.1\textwidth]{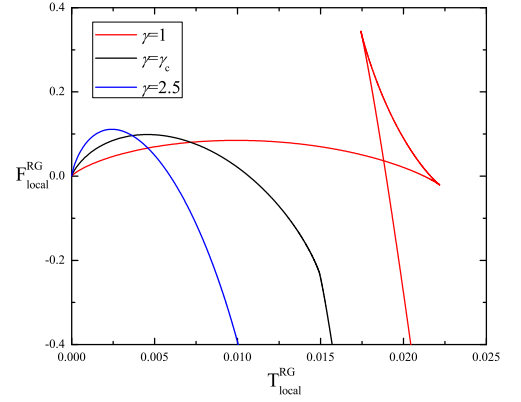}
\label{fig3-a}
\end{minipage}
}
\subfigure[]{
\begin{minipage}[b]{0.41\textwidth}
\includegraphics[width=1.1\textwidth]{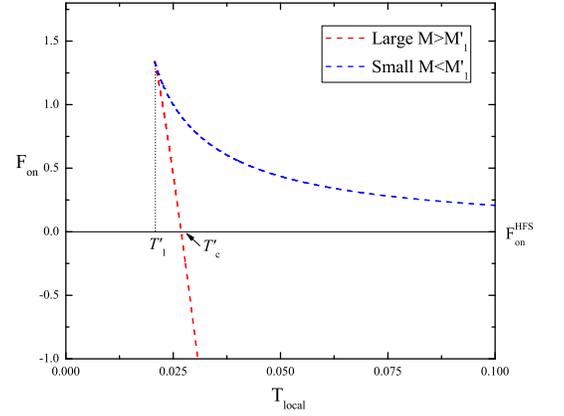}
\label{fig3-b}
\end{minipage}
}
\subfigure[]{
\begin{minipage}[b]{0.41\textwidth}
\includegraphics[width=1.1\textwidth]{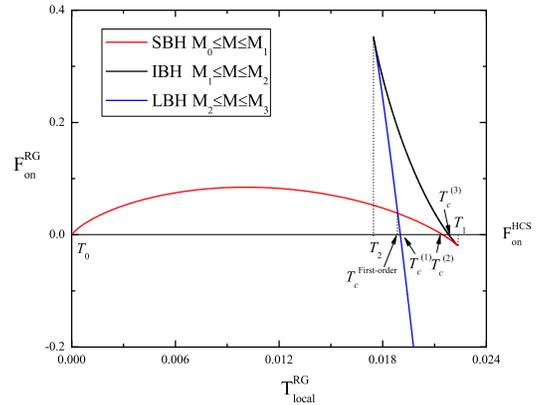}
\label{fig3-c}
\end{minipage}
}
\caption{(a) Variation in free energy of rainbow SC black hole with local temperature for different $\gamma$.
(b) and (c) Original free energy of SC black hole and rainbow SC black hole as a function of local temperature. Here, we choose $G = 1$ and $r=10$.}
\label{fig3}       % Give a unique label
\end{figure}

In order to obtain more details on the thermodynamic phase transition and stability of the rainbow SC black hole, it is necessary to investigate the free energy of rainbow SC black hole in a cavity. In \cite{ch67,ch68}, the free energy is defined as follows:
\begin{equation}
\label{eq12+}
F_{\rm{on}}  = E_{\rm{local}}  - T_{\rm{local}} S.
\end{equation}
Substituting Eq.~(\ref{eq9}), Eq.~(\ref{eq10}) and Eq.~(\ref{eq11}) into Eq.~(\ref{eq12+}), the rainbow free energy is given by
\begin{align}
\label{eq13}
F_{\rm{on}}^{\rm{RG}}& =  \frac{r}{G}\left( {\sqrt {1 - \frac{{\gamma G^{{3 \mathord{\left/ {\vphantom {3 2}} \right. \kern-\nulldelimiterspace} 2}} }}{r}}
 - \sqrt {1 - \frac{{2GM}}{r}} } \right)
\nonumber \\
& - \frac{{\left( {2M - \sqrt G \gamma } \right)\chi}}{{8M^2 \sqrt {1 - \frac{{2GM}}{r}} }},
\end{align}
where $\chi=  {2M\left( {M + \sqrt G \gamma } \right)  + G\gamma ^2 {\rm{ln}}\left( {{{2M} \mathord{\left/ {\vphantom {{2M} {\sqrt G }}} \right. \kern-\nulldelimiterspace}  {\sqrt G }} - \gamma } \right)}$. When $\gamma=0$, Eq.~(\ref{eq13}) is reduced to the original free energy  $ F_{\rm{on}}  = r{{\left( {1 - \sqrt {1 - {{2GM} \mathord{\left/ {\vphantom {{2GM} r}} \right. \kern-\nulldelimiterspace} r}} } \right)} \mathord{\left/ {\vphantom {{\left( {1 - \sqrt {1 - {{2GM} \mathord{\left/ {\vphantom {{2GM} r}} \right. \kern-\nulldelimiterspace} r}} } \right)} G}} \right. \kern-\nulldelimiterspace} G} - {M \mathord{\left/ {\vphantom {M {2\sqrt {1 - {{2GM} \mathord{\left/ {\vphantom {{2GM} r}} \right. \kern-\nulldelimiterspace} r}} }}} \right. \kern-\nulldelimiterspace} {2\sqrt {1 - {{2GM} \mathord{\left/ {\vphantom {{2GM} r}} \right. \kern-\nulldelimiterspace} r}} }}$. For further investigation of the phase transition of rainbow SC black hole, we plot Fig.~\ref{fig3}.

Fig.~\ref{fig3-a} shows $F_{\rm{on}}^{\rm RG}$ curves with $T_{\rm{local}}^{\rm RG}$ for different $\gamma$, the value of rainbow parameter $\gamma$ decreases from top to bottom. The swallow tail structure appears when the rainbow parameter $\gamma$ is smaller than the critical value $\gamma_{c}$, which indicates that there is a two-phase coexistence state. The results are consistent with the profile of $T_{\rm{local}}^{\rm RG}-M$ in Fig.~\ref{fig1-a}.

Fig.~\ref{fig3-b} and Fig.~\ref{fig3-c} show the free energy of original SC black hole and rainbow SC black hole, respectively. First, let us focus on the original case for $\gamma=0$. It is found that the local temperature has a minimum value $T_1'$, below which no black hole solution exists. Moreover, the free energy vanishes when $M \to {\rm{0}}$  because the vacuum state is Minkowski spacetime. Hence, $F_{\rm{on}}^{{\rm{HFS}}}$ represents the free energy of the hot flat space (HFS).  The Hawking-Page transition occurs at $T_c'$. For  $T < T_c ^\prime$, the free energies of both unstable small black hole and stable large black hole are higher than $F_{\rm{on}}^{{\rm{HFS}}}$, this implies that the HFS is more probable than the unstable small black hole and stable large black hole. However, when  $T > T_c ^\prime$, the free energy of unstable small black hole is higher than $F_{\rm{on}}^{{\rm{HFS}}}$, while the free energy of stable large black hole is lower than $F_{\rm{on}}^{{\rm{HFS}}}$. This indicates that the stable large black hole is more probable than the HFS. Therefore, one can see that the radiation collapse to the stable large black hole by phase transition, and the unstable small black hole eventually decays into the large black hole thermodynamically for $T > T_c'$.

Then, turn to look at Fig.~\ref{fig3-c}. For non-zero  $\gamma$, the behavior of the rainbow free energy is different from that of the original case:

(i)  Due to the presence of black hole remnants, we need use the hot curved space (HCS) to investigate phase transition of rainbow SC black hole instead of using the HFS \cite{ch60+}. Therefore, we analyse the phase transition between the three black hole states and the HCS via studying free energies of rainbow SC black hole. In addation, the HCS becomes zero when $M\rightarrow M_{\rm res}$, i.e., $F_{\rm{on}}^{{\rm{HCS}}}=0$.

(ii) The small-intermediate black hole transition occurs at the inflection point $T_1$  corresponding to the mass  $M_1$, the IBH and LBH meet at $T_2$  corresponding to the mass $M_2$. Interestingly, three Hawking-Page-type critical points can be found easily, which correspond to the temperatures $T_{c}^{(1)}$, $T_{c}^{(2)}$, and $T_{c}^{(3)}$, respectively, as shown in Fig.~\ref{fig3-c}, whereas only one Hawking-Page phase transition point can be found in original case, as shown in Fig.~\ref{fig3-a}.

(iii) A first-order phase transition occurs because the $F_{\rm{on}}$ demonstrates the characteristic swallow tail behavior. It is reminiscent of the ``free energy-Hawking temperature'' relation of charged AdS black hole, as described in \cite{ch51,ch53,ch54,ch55,ch56,ch57}. The intersection point between the red solid line and the blue solid line is the first-order phase transition point that correspond to the temperature $T_c^{\rm First-order}$.

(iv) For $T_0 < T < T_{c}^{(1)}$, the free energies of the three black holes are higher than those of the HCS. Therefore, the HCS is more probable in this region. Meanwhile, it can be observed that the IBH and LBH collapse into the SBH for $T_0 < T < T_{c}^{\rm First-order}$ because $F_{\rm{on}}^{{\rm{SBH}}}< F_{\rm{on}}^{{\rm{LBH}}} < F_{\rm{on}}^{{\rm{IBH}}}$, and the IBH and SBH collapse into the LBH for $T_{c}^{\rm First-order} < T < T_{c}^{\rm (1)}$ since $F_{\rm{on}}^{{\rm{LBH}}}< F_{\rm{on}}^{{\rm{SBH}}} < F_{\rm{on}}^{{\rm{IBH}}}$. However, the free energies of the LBH, SBH and IBH decrease below $ F_{\rm{on}}^{\rm HCS}$ one by one when the temperature is higher than $T_{c}^{(1)}$. For $T_{c}^{(1)} < T < T_{c}^{(2)}$, the relation of free energies obey $F_{\rm{on}}^{{\rm{LBH}}} < F_{\rm{on}}^{{\rm{HCS}}} < F_{\rm{on}}^{{\rm{SBH}}} < F_{\rm{on}}^{{\rm{IBH}}} $. It then changes to $F_{\rm{on}}^{{\rm{LBH}}} < F_{\rm{on}}^{{\rm{SBH}}}  < F_{\rm{on}}^{{\rm{HCS}}} < F_{\rm{on}}^{{\rm{IBH}}} $ for $T_{c}^{(2)} < T < T_{c}^{(3)}$. Finally, the relation becomes $F_{\rm{on}}^{{\rm{LBH}}} < F_{\rm{on}}^{{\rm{SBH}}} < F_{\rm{on}}^{{\rm{IBH}}} < F_{\rm{on}}^{{\rm{HCS}}}$ for $T_{c}^{(3)} < T < T_1$. Therefore, the stable SBH and unstable IBH decay into the stable LBH eventually for $T_{c}^{(1)} < T < T_1$.

\section{Discussion}
\label{Dis}
In this paper, using the rainbow functions that were proposed by Magueijo and Smolin, we studied the quantum corrections to the thermodynamics and the phase transitions of SC black hole. According to the rainbow surface gravity and the uncertainty principle, we calculated modified thermodynamic quantities such as the rainbow Hawking temperature, rainbow entropy, rainbow heat capacity, and rainbow local temperature. From these modifications, we then derived the critical points of the black hole thermodynamic ensemble, and analyzed the thermodynamic stability and phase transition of the rainbow SC black hole.

Comparing our results with the original thermodynamic quantities and Hawking-Page phase transition of SC black hole, many different findings were obtained. First, the rainbow gravity can stop the Hawking radiation in the final stages of black holes' evolution and lead to remnants, which is in agreement with the predictions of the generalized uncertainty principle. Second, from Eq.~(\ref{eq12}), one can obtain two second-order phase transitions because the heat capacity exhibits two divergences: at $M_1$ and $M_2$. Third, according to the ``$F_{\rm{on}}^{\rm RG}-T_{\rm{local}}^{\rm RG}$'' plane, it was found that an unstable  intermediate black hole, which never appears in the original SC black hole, interpolates between the stable small black hole and stable large  black hole in the rainbow SC black hole. Fourth, $F_{\rm{on}}^{\rm RG}$ surface demonstrated the characteristic swallow tail behavior, which implies that the rainbow SC black hole system has a first-order transition. This behavior is reminiscent of the ``free energy-Hawking temperature'' relation in AdS spacetimes. Finally, three Hawking-Page-type critical points were found for rainbow SC black hole, as shown in Fig.~\ref{fig3-c}, whereas only one Hawking-Page critical point was found for the original case (see Fig.~\ref{fig3-b}) and two Hawking-Page critical points were observed in~\cite{ch59}. For $T_0 < T < T_{c}^{(1)}$, the IBH and LBH will collapse into the stable SBH since the free energy of SBH is lower than those of the IBH and LBH. However, for $T_0 < T < T_{c}^{(1)}$, the SBH and IBH decay into the stable LBH because the free energy of LBH is lower than those of the IBH and SBH.

\section*{Acknowledgements}
The authors thank Yongwan Gim and the anonymous referees for helpful suggestions and enlightening comments, which helped to improve the
quality of this paper. This work was  supported by the Natural Science Foundation of China (Grant No. 11573022).

\end{document}